\documentclass[sigconf, screen]{acmart}

\usepackage{enumitem}

\AtBeginDocument{%
  }

\copyrightyear{2025}
\acmYear{2025}
\setcopyright{rightsretained}
\acmConference[AutomotiveUI Adjunct '25]{17th International Conference on Automotive User Interfaces and Interactive Vehicular Applications}{September 21--25, 2025}{Brisbane, QLD, Australia}
\acmBooktitle{17th International Conference on Automotive User Interfaces and Interactive Vehicular Applications (AutomotiveUI Adjunct '25), September 21--25, 2025, Brisbane, QLD, Australia}
\acmDOI{10.1145/3744335.3758495}
\acmISBN{979-8-4007-2014-7/2025/09}

\begin{document}





\title[Understanding Pedestrian Gesture Misrecognition via VLM Reasoning]{Understanding Pedestrian Gesture Misrecognition: Insights from Vision-Language Model Reasoning}

\author{Tram Thi Minh Tran}
\email{tram.tran@sydney.edu.au}
\orcid{0000-0002-4958-2465}
\affiliation{Design Lab, Sydney School of Architecture, Design and Planning,
  \institution{The University of Sydney}
  \city{Sydney}
  \state{NSW}
  \country{Australia}
}

\author{Xinyan Yu}
\email{xinyan.yu@sydney.edu.au}
\orcid{0000-0001-8299-3381}
\affiliation{Design Lab, Sydney School of Architecture, Design and Planning,
  \institution{The University of Sydney}
  \city{Sydney}
  \state{NSW}
  \country{Australia}
}

\author{Callum Parker}
\email{callum.parker@sydney.edu.au}
\orcid{0000-0002-2173-9213}
\affiliation{Design Lab, Sydney School of Architecture, Design and Planning,
  \institution{The University of Sydney}
  \city{Sydney}
  \state{NSW}
  \country{Australia}
}

\author{Julie Stephany Berrio Perez}
\email{stephany.berrioperez@sydney.edu.au}
\orcid{0000-0003-3126-7042}
\affiliation{The Australian Centre for Robotics,
  \institution{The University of Sydney}
  \city{Sydney}
  \state{NSW}
  \country{Australia}
}

\author{Stewart Worrall}
\email{stewart.worrall@sydney.edu.au}
\orcid{0000-0001-7940-4742}
\affiliation{The Australian Centre for Robotics,
  \institution{The University of Sydney}
  \city{Sydney}
  \state{NSW}
  \country{Australia}
}

\author{Martin Tomitsch}
\email{martin.tomitsch@uts.edu.au}
\orcid{0000-0003-1998-2975}
\affiliation{Transdisciplinary School,
  \institution{University of Technology Sydney}
  \city{Sydney}
  \state{NSW}
  \country{Australia}
}

\renewcommand{\shortauthors}{Tran et al.}




\begin{abstract} 



Pedestrian gestures play an important role in traffic communication, particularly in interactions with autonomous vehicles (AVs), yet their subtle, ambiguous, and context-dependent nature poses persistent challenges for machine interpretation. This study investigates these challenges by using GPT-4V, a vision–language model, not as a performance benchmark but as a diagnostic tool to reveal patterns and causes of gesture misrecognition. We analysed a public dataset of pedestrian–vehicle interactions, combining manual video review with thematic analysis of the model’s qualitative reasoning. This dual approach surfaced recurring factors influencing misrecognition, including gesture visibility, pedestrian behaviour, interaction context, and environmental conditions. The findings suggest practical considerations for gesture design, including the value of salience and contextual redundancy, and highlight opportunities to improve AV recognition systems through richer context modelling and uncertainty-aware interpretations. While centred on AV–pedestrian interaction, the method and insights are applicable to other domains where machines interpret human gestures, such as wearable AR and assistive technologies.

\end{abstract}



\maketitle


\section{Introduction}

Pedestrian gestures, such as hand waves, play an important role in traffic communication~\cite{rasouli2018social, stanciu2018pedestrians, tran2024mapping}, even though they occur infrequently~\cite{lee2021road}. Gestures here refer to deliberate bodily movements intended to convey intent~\cite{ekman2004emotional, ekman1969repertoire}. As autonomous vehicles (AVs)~\cite{rasouli2018social, wang2022shared} and urban robots~\cite{yu2024understanding} take on similar roles, their ability to accurately interpret these non-verbal signals is essential for seamless integration into dynamic and socially complex environments~\cite{wang2022shared}. However, the recognition and interpretation of pedestrian gestures remain highly context-dependent. For example, in a large-scale study on pedestrian-driver communication, \citet{rasouli2018social} found that not all hand movements are intentional gestures directed at vehicles. Pedestrians may adjust their clothing or raise their hands for reasons unrelated to traffic, introducing ambiguities that challenge gesture recognition in these systems. Despite this, there has been limited investigation into the specific factors that lead recognition systems to misinterpret gestures. Most prior work has prioritised improving overall classification accuracy, offering less insight into why such errors occur. Understanding these causes is critical for informing both gesture design and system development.


To address this gap, we investigate how a vision–language model (VLM) interprets pedestrian gestures in real-world traffic scenes, focusing on cases where recognition fails or is uncertain. Rather than evaluating performance, we aim to identify recurring factors that contribute to misinterpretation. VLMs are increasingly explored for AV perception tasks, offering a way to combine visual and contextual understanding in complex environments~\cite{huang2024gpt, wandelt2024large}. Among them, GPT-4 Vision (GPT-4V) exemplifies this potential, showing qualitative proficiency in interpreting pedestrian behaviour and scene dynamics~\cite{driessen2024putting}. Building on these capabilities, we use GPT-4V to analyse a subset of videos from the Joint Attention in Autonomous Driving (JAAD) dataset~\cite{rasouli2017they}, focusing on misrecognised gestures and the model's accompanying reasoning. This approach allows us to surface influential factors, including gesture visibility, pedestrian behaviour, interaction context, and environmental conditions.

\textit{Contribution Statement}: 
This study identifies key factors that influence how VLMs interpret pedestrian gestures, based on cases of misrecognition. While our focus is on AV–pedestrian interactions, the findings offer broader insights into the challenges of gesture interpretation in real-world contexts. These insights may inform future work in gesture design, standardisation, and system development across domains where machines interpret human gestures, including urban robotics, wearable AR interfaces, and assistive technologies. More broadly, the study demonstrates how VLM reasoning can be used as a diagnostic tool to expose context-dependent recognition issues that may be overlooked in performance-focused evaluations.







\section{Related Work}

VLMs, such as GPT-4V, are gaining attention in research on advanced mobility systems. In human-robot interaction, these models have been used to generate captions for in-the-wild datasets~\cite{bu2024restory}. In AV research, they have shown promise in tasks such as predicting pedestrian behaviour~\cite{huang2024gpt}, assessing risks~\cite{driessen2024putting}, and predicting hazards~\cite{charoenpitaks2024exploring} in traffic environments. Other studies, including \citet{onkhar2024towards}, have explored how GPT-4V integrates multimodal data to improve context-sensitive safety systems, while \citet{pan2024vlp} examined its application in vision language planning for autonomous driving, focusing on connecting visual input with driving decisions.

Despite these advancements, gesture recognition remains a particular challenge. \citet{bossen2025can} found that while some state-of-the-art VLMs (e.g., VideoLLaMA2, VideoLLaMA3, and Qwen2\footnote{ChatGPT-4o was excluded from this study as the authors limited their assessment to models that can be run locally on an edge device without internet access or API charges.}) can interpret human traffic gestures in zero-shot settings, their outputs are neither accurate nor robust enough to be considered reliable. These limitations reflect broader challenges in AV–pedestrian interaction, where distinguishing intentional gestures from incidental movements remains difficult~\cite{stanciu2018pedestrians, rasouli2018social}. The standardisation of pedestrian gestures has been proposed as one strategy to improve recognition reliability~\cite{stanciu2018pedestrians}, typically by identifying gestures~\cite{tran2024mapping, brand2025identifying} that are intuitive for pedestrians and culturally appropriate. However, recognition accuracy may also be shaped by how gestures are executed or by contextual conditions, factors that remain underexplored.

While current VLMs~\cite{bossen2025can} may struggle with consistent gesture recognition, recent studies have highlighted their potential for exploratory reasoning and qualitative interpretation of traffic scenes~\cite{huang2024gpt, driessen2024putting, onkhar2024towards}. For example, \citet{huang2024gpt} demonstrated that GPT-4V can describe pedestrian behaviour, interpret group dynamics, and incorporate contextual cues when explaining its decisions. Building on this interpretive capacity, our study examines how GPT-4V responds to ambiguous or misclassified pedestrian gestures. By analysing not only its recognition outcomes but also its accompanying reasoning, we aim to surface factors that contribute to misrecognition, insights that may not be readily accessible through traditional computer vision approaches.





\section{Method}

\begin{figure*}[h!]
    \centering
    \includegraphics[width=1\linewidth]{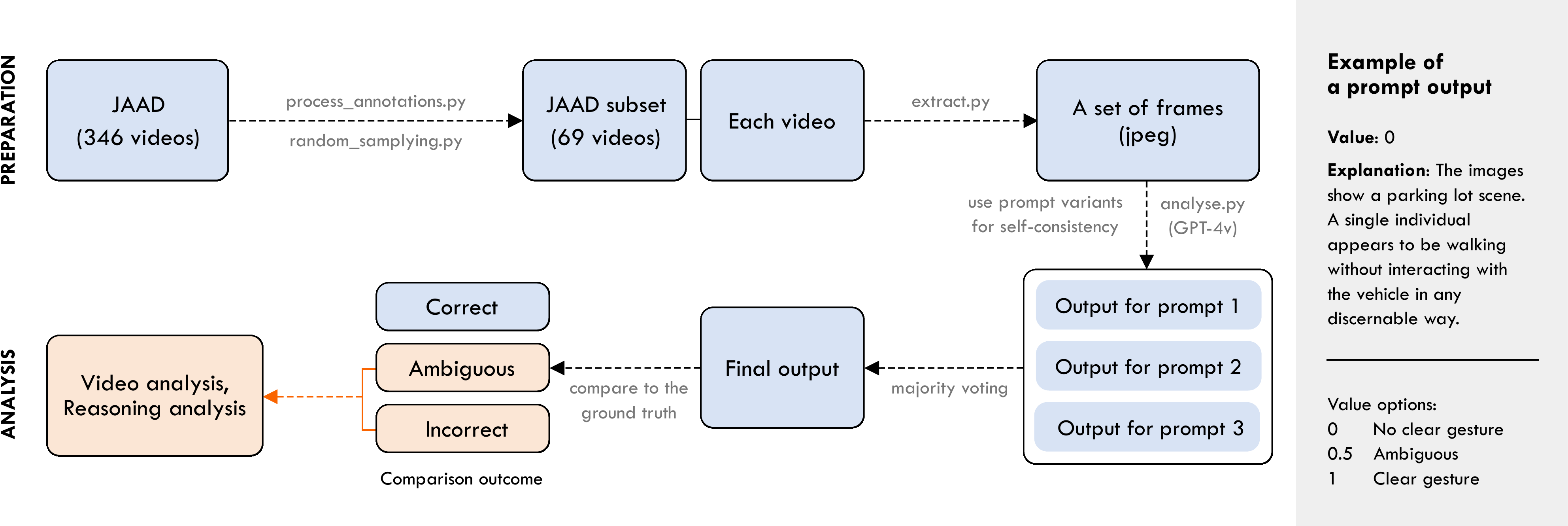}
    \caption{Workflow for analysing gesture recognition in the JAAD dataset.}
    \Description{Workflow for analysing gesture recognition in the JAAD dataset.}
    \label{fig:workflow}
\end{figure*}

\autoref{fig:workflow} illustrates the overall workflow used to analyse gesture recognition in the JAAD dataset, structured into two phases: Video Preparation and Analysis. All scripts used in this workflow are available in \autoref{sec:scripts}.

\subsection{Video Preparation}

\subsubsection{Selecting Videos}

This study used the JAAD dataset~\cite{rasouli2017they}, which consists of 346 video recordings, each 5–10 seconds long, capturing naturalistic pedestrian-vehicle interactions in uncontrolled environments. The dataset includes XML annotation files~\cite{JAAD_annotations} that label pedestrian behaviours and actions for each video. Using these annotations, we identified only 23 out of 346 videos containing explicit gestures directed at the vehicle. To optimise computational resources and focus the analysis, we selected a subset of videos comprising three categories:
\begin{itemize}
\item \textit{gestures}: videos with explicit gestures.
\item \textit{looking-only}: videos where pedestrians look at the vehicle but exhibit no gestures.
\item \textit{none}: videos where pedestrians neither look at nor gesture toward the vehicle.
\end{itemize}
Since only 23 videos contained explicit gestures, we selected 23 videos for each of the other two categories to ensure a balanced representation of gesture presence, ambiguous cases, and neutral interactions. A Python script was used to process the annotation files and randomly sample 23 videos from the remaining dataset for each of the \textit{looking-only} and \textit{none} categories.


\subsubsection{Converting to Images}

As GPT-4V cannot directly analyse video and instead processes base64-encoded images, OpenCV was used to extract evenly spaced frames from each video. For the \textit{gestures} subset, 50 frames were extracted to ensure detailed temporal coverage of rapid and subtle actions. For the \textit{looking-only} and \textit{none} subsets, 10 frames were extracted to provide a representative sampling of the content while maintaining computational efficiency~\cite{huang2024gpt}.

\subsection{GPT-4V Analysis}

We used GPT-4o, OpenAI's flagship model\footnote{\url{https://platform.openai.com/docs/models/gpt-4o}}, to analyse the recognition of pedestrian gestures in AV-pedestrian interactions. The specific model used was \texttt{chatgpt-4o-latest}, which continuously updates to the latest version of GPT-4o available in ChatGPT.  

\subsubsection{Prompt Engineering}

Language models are highly sensitive to the way questions are posed. To address this, we followed OpenAI's prompt engineering guide to design effective prompts\footnote{ \url{https://platform.openai.com/docs/guides/prompt-engineering}}
The selected strategy involved crafting a System message to clearly define the model's task scope and expected outputs. For the User message, we applied self-consistency~\cite{driessen2024putting} by using multiple prompt variations. Specifically, we iteratively tested and selected three variations of the User message—Generic, Clarity, and Factors—each emphasising different aspects of gesture recognition (see \autoref{tab:prompt_variation}). To streamline prompting for each video, we developed a Python script that automatically applied the prompts to the dataset. All prompting was conducted on 16 January 2025.

\begin{table}[h!]
\caption{A basic prompt structure consisting of System and User messages, with User message variations to explore self-consistency.}
\renewcommand{\arraystretch}{1.5} 
\centering
\small
\begin{tabular}{p{0.1\linewidth}p{0.1\linewidth}p{0.7\linewidth}} 
\textbf{SYSTEM} & \multicolumn{2}{p{0.83\linewidth}}{
GPT-4 computer vision for images: enabled. You have the capability to analyse multiple images and provide a combined description. Assign a single numeric value for the sequence:
- 0 if no gestures are observed or all gestures are unrelated to the vehicle,
- 0.5: if there is some evidence of a gesture (excluding walking), but its intent is uncertain or the gesture is only partially visible.
- 1 if clear gestures directed toward the vehicle are observed.
Provide reasoning for the assigned value in your description.}\\
\textbf{USER} & [Generic] & Based on these images, provide a numeric value (0, 0.5, or 1). \textit{Briefly describe the observed gestures and their intent toward the vehicle.}\\
& [Clarity] & Based on these images, provide a numeric value (0, 0.5, or 1). \textit{Explain why the gesture is (or is not) intentional and directed at the vehicle, and whether it appears visually clear or subtle.}\\
& [Factors] & Based on these images, provide a numeric value (0, 0.5, or 1). \textit{Highlight any factors (e.g., environmental conditions, social or spatial dynamics) that might influence the clarity or misrecognition of these gestures.}\\
\end{tabular}
\label{tab:prompt_variation}
\end{table}






\subsubsection{Flagging Incorrect and Ambiguous Cases}

After applying the prompt to all videos, we aggregated the outputs from multiple prompt variations for each video using majority voting. For example, if the outputs are [0,~0.5,~0], the result is 0, indicating that no gesture was observed. In evenly distributed outputs, [0,~0.5,~1], the result defaults to 0.5. The aggregated result was compared to the binary ground truth (0 or 1). Videos where the aggregated result does not match the ground truth were flagged as \textit{incorrect}, and videos with an aggregated result of 0.5 were flagged as \textit{ambiguous}.

\subsubsection{Analysis of Recognition Challenges}

This process involved two key components: manual video analysis and model reasoning analysis. For manual video analysis, the first author reviewed each flagged video to describe the interaction scenarios and gestures observed. These manual descriptions went beyond the annotation files provided by JAAD, allowing us to identify potential factors contributing to recognition errors based on human observations.

The model reasoning analysis focused on the explanations provided by the GPT-4V through its prompt outputs. We thematically analysed the model’s reasoning to understand how it came to its decision. This process was reflexive, involving two independent coders who later engaged in discussions to explore various perspectives. 



\section{Results}

\subsection{Gesture Recognition Outcome}

\begin{figure*}[h!]
    \centering
    \includegraphics[width=0.55\linewidth]{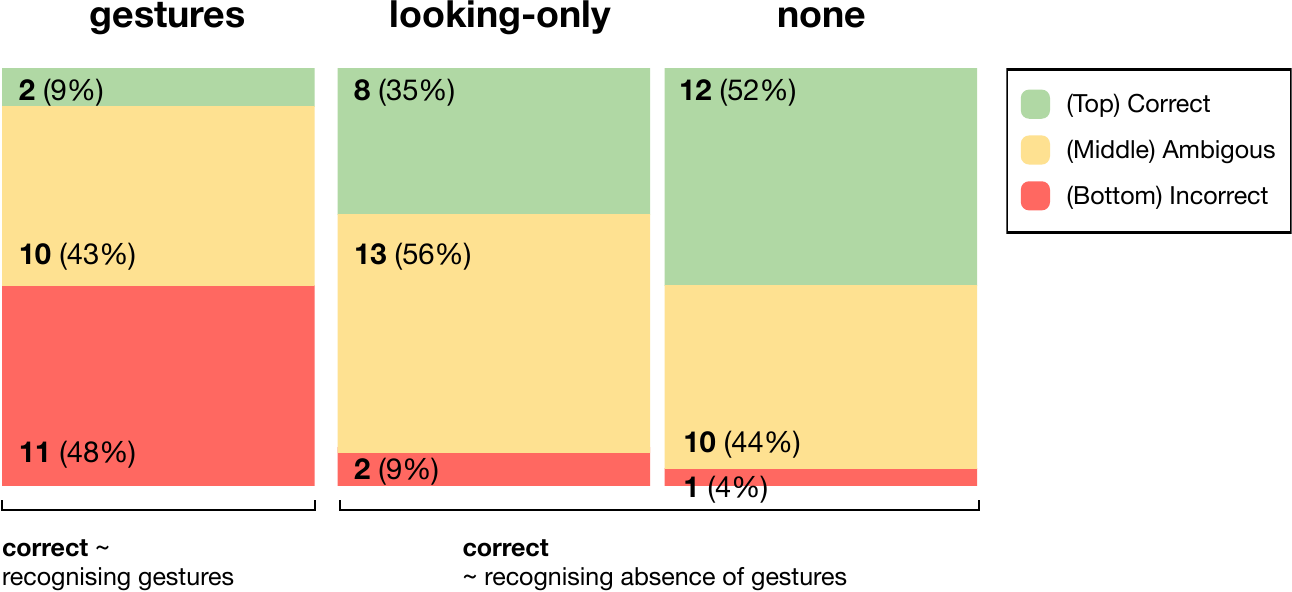}
    \caption{Summary of gesture recognition outcome across categories. `Correct' (green) refers to recognising gestures in the \textit{gestures} category and recognising the absence of gestures in the \textit{looking-only} and \textit{none} categories.}
    \Description{Summary of gesture recognition outcome across categories. `Correct' (green) refers to recognising gestures in the \textit{gestures} category and recognising the absence of gestures in the \textit{looking-only} and \textit{none} categories.}
    \label{fig:outcome}
\end{figure*}

\autoref{fig:outcome} highlights distinct trends across the categories. The \textit{gestures} category shows significant challenges, with a low proportion of correct classifications and high incorrect rates. In contrast, the \textit{looking-only} and \textit{none} categories demonstrate stronger performance, characterised by higher correct rates and minimal incorrect classifications. Ambiguity remains a consistent outcome across all categories.

\subsection{Manual Video Analysis}

In the 21 flagged videos from the \textit{gestures} category, gestures that were not recognised or considered ambiguous by the model were generally subtle, quick, and low in prominence (n=16). 
Many subtle gestures, such as hand-raising or waving, performed near the waist or chest height, blend easily with clothing or natural body movements. Some gestures involved holding objects like shopping bags or purses, occasionally resembling unrelated actions, such as adjusting a shoulder bag. Nods were deliberate but brief, often requiring focused attention to notice. One gesture, annotated as a nod, could not be verified by the researcher even after repeated observations. Few videos featured slightly more explicit gestures (n=5), such as hand-raising above shoulder height, a nod resembling a bow, or open palms with fingers spread, which were more visually distinct.

In the 26 flagged videos from the \textit{looking-only} and \textit{none} categories, incorrect and ambiguous cases arose from three primary types of pedestrian actions: unrelated actions while waiting or crossing, such as holding a phone or adjusting a hat (n=4); individuals near crosswalks or roadsides who looked at the vehicle before or during the crossing (n=12); and pedestrians already crossing when the vehicle was far away, nearing completion as the vehicle arrived (n=10).

Across categories, gestures or actions occurred at varying distances from the vehicle: close proximity (1–3 m), moderate distances (3–5 m) and far distances (6–10 m). These actions were observed in various environmental contexts, including marked and unmarked crossings, outdoor parking lots, weather conditions ranging from overcast skies to clear daylight, and scene complexities involving single pedestrians or dynamic environments with multiple actors. There was no single external factor consistently led to incorrect or ambiguous outputs. 

\subsection{Model Reasoning Analysis}

\subsubsection{How Gestures Were Recognised?}
\label{sec:how-reasoning}

The model's reasoning process involved evaluating key aspects of gestures while considering additional factors such as pedestrian behaviour, interaction contexts, and video quality to determine whether a pedestrian gesture existed and was directed at the vehicle.

\textbf{\textcolor{teal}{Gestures.}}
The model determined whether an action qualified as a gesture primarily based on \textbf{hand or arm movements} such as \textit{`waving,'} \textit{`pointing,'} \textit{`stopping motions,'} or \textit{`any other distinguishable or distinctly recognisable gesture.'} It also evaluated the gesture's \textbf{visibility and explicitness}, determining whether it appeared \textit{`explicit,'} \textit{`sustained,'} \textit{`overt,'} \textit{`dominant',} \textit{`exaggerated,'} or \textit{`outward.'} Gestures with subtle characteristics that are \textit{`partially visible,`} \textit{`not visually clear enough,'} often made it difficult for the model to classify them as deliberate. For example, a slight arm raise could be interpreted as incidental or unrelated, such as maintaining balance or gesturing playfully.

\textbf{\textcolor{teal}{Pedestrian Behaviour.}}
The model relied on pedestrian-specific cues to determine if there was a potential interaction \textit{with} the vehicle. \textbf{Directionality}, expressed through direct eye contact or body orientation toward the vehicle, often indicated intent. For instance, a pedestrian \textit{`turns her face toward the vehicle, potentially indicating awareness of it,'} or \textit{`seemingly waves while looking toward the car.'} \textbf{Position} also played a role; pedestrians near the vehicle or standing at the edge of a crosswalk were more likely to be associated with interaction intent. For example, \textit{`distances between the vehicle and the individuals conveying an unclear sense of proximity or need for interaction.'} The model further considered \textbf{trajectory}, i.e., the pedestrian’s movement relative to the vehicle’s path, as an important cue. \textit{`Walking motion into and across the vehicle's lane'} or \textit{`crossing [the vehicle's] path'} often indicated a clear intent to interact. However, if pedestrian \textbf{movements} were described as \textit{`natural,'} \textit{`consistent,'} \textit{`routine,'} \textit{`continuous,'} or \textit{`normal'}, they were often interpreted as standard walking or crossing behaviours, primarily focused on navigating their environment rather than attempting to interact. \textbf{Carrying objects or multitasking} introduced ambiguity. For example, \textit{`The man near the car holding an item (possibly a phone). His actions do not clearly manifest as directed specifically toward the vehicle.'} 

\textbf{\textcolor{teal}{Interaction Context.}}
The interaction contexts influenced the model’s reasoning process, either adding or subtracting uncertainty from its decision-making. \textbf{Designated crossing areas} often reduced the reliance on gestures to signal intent, as the context implied a right-of-way. For example, signage such as `Watch for Pedestrians' and marked crosswalks naturally established expectations for pedestrian behaviour, minimising the need for explicit gestures directed at vehicles. \textbf{Scene complexity} also played a critical role in the model’s reasoning. In scenarios involving multiple pedestrians and vehicles, distractions or competing focus points diluted the recognition of individual gestures. The presence of other pedestrians \textit{`might confound the determination of whether gestures are directed at the vehicle.'} A group of pedestrians crossing together \textit{`may lead drivers to assume right-of-way, even without explicit gestures.'}

\textbf{\textcolor{teal}{Video Quality.}}
The model’s reasoning assessed the degree to which video quality factors obstructed gesture recognition. Among these, \textbf{weather conditions}, such as \textit{`snowfall,'} \textit{`rain,'} \textit{`droplets on the windshield,'} \textit{`overcast skies,'} \textit{`sun glares,'} \textit{`light reflections,'} and \textit{`shadows,'} were frequently noted. The \textbf{camera angle} also played a role, particularly in cases where \textit{`the field of view limits clear observation of smaller or more subtle movements'} or \textit{`moments where people appear closer to the edges of the frame might reduce clarity of perception.'} Additionally, \textbf{vehicle speed} influenced the time available for observation. For instance, slower-moving vehicles provided \textit{`sufficient time to observe the pedestrian and interpret their actions.'}

\subsubsection{What Caused Misrecognition?}

In the \textit{gestures} category, recognition failures or ambiguities were primarily due to subtle or low-visibility gestures, as reflected in repeated model reasoning: \textit{`no visible hand motions, waving, or pointing directed at the vehicle'} (n=21). Additional factors related to pedestrian behaviour, interaction context, and video quality further contributed to the ambiguity (n=7). 

In the \textit{looking-only} and \textit{none} categories, false positives and ambiguities were driven by noticeable arm movements that the model misinterpreted as gestures (n=4). Pedestrian cues, such as directionality, proximity to the vehicle, and deliberate crossing behaviour, often led the model to infer an interaction even when no gesture was present (n=24). 

\section{Discussion}

\subsection{What Misrecognition Reveals About Gesture Interpretation}

Previous research, such as that by \citet{rasouli2018social}, underscores key challenges in pedestrian-vehicle communication relevant to gesture recognition systems. They highlight the importance of identifying relevant interactions, making sense of gestures within their context, and distinguishing symbolic gestures from incidental movements. Our study corroborates these challenges and extends their work by examining how a VLM (GPT-4V) processes pedestrian gestures, identifying key factors that contribute to misrecognition and ambiguity.

GPT-4V’s reasoning frequently pointed to visibility and explicitness as key factors affecting recognition. Our manual video analysis arrived at the same conclusion: brief and subtle gestures posed significant recognition challenges. These issues were influenced by the execution of gestures and video quality, making certain movements harder for GPT-4V to detect or correctly interpret. Beyond detecting noticeable movement (e.g., a hand raise), GPT-4V’s reasoning highlighted the challenge of interpreting a gesture’s intent (intentional vs. incidental) and its intended target (directed at the AV vs. unrelated). This second layer of interpretation depends on pedestrian behaviour and interaction context. Rather than drawing prescriptive design rules, we interpret these findings as indicators of recurring points of friction. Below, we summarise several practical considerations that emerged across misrecognised cases:
\begin{itemize}
    \item \textbf{Salience matters}. Gestures that are large, deliberate, and sustained are more reliably interpreted by the model. Movements that are subtle or brief, especially those partially occluded or performed in motion, are often overlooked.
    \item \textbf{Contextual redundancy helps}. Combining gestures with body orientation (e.g., facing the vehicle), pausing before crossing, or making eye contact can reinforce crossing intent.
\end{itemize}

These patterns suggest that gesture recognition challenges arise not only from current system limitations, but also from the inherently fluid, variable, and often ambiguous nature of pedestrian gestures in real-world contexts. As such, efforts to support AV–pedestrian interaction should be cautious about overly relying on gesture standardisation. Instead, recognition systems may need to incorporate richer context modelling and uncertainty-aware interpretations.


\subsection{Strengths and Limitations of VLM-Based Reasoning}

Reasoning analysis showed that GPT-4V provides strong scene understanding, aligning with findings from ~\cite{huang2024gpt}. It offers valuable insights into factors influencing gesture recognition and helps uncover potential biases that shape model assumptions, informing adjustments to training data. For example, contextual cues like crosswalks or signage can bias models to assume explicit gestures are unnecessary, making them less likely to detect courtesy or acknowledgment gestures, such as small nods or hand waves~\cite{tran2024mapping}. 

However, while GPT-4V can identify many factors influencing gesture recognition, they cannot weigh their relative importance. This limitation can lead to inconsistencies between the model’s reasoning and its recognition outcome. In many cases, the model detected the absence of explicit hand gestures but still classified the scenario as ambiguous due to contextual cues suggesting interaction (e.g., \autoref{fig:example} right). Furthermore, it is crucial to recognise that the reasoning provided by VLMs like GPT-4V relies on textual analysis of images, which may not always align with the actual visual content, an issue known as hallucination~\cite{liu2024survey}. This can misattribute causes or infer intent beyond what is explicitly present in the scene. The reliability of VLM reasoning also depends on the model’s sophistication, and the GPT-4V model used in our study exhibited common issues identified in prior research~\cite{huang2024gpt, bossen2025can}, including prompt dependence, inconsistent outputs, and difficulty comprehending complex scenes. As the model evolves with updates or enhancements, its capabilities and limitations may shift.

Despite these shortcomings, VLMs can serve not only as recognition tools but also as diagnostic lenses for analysing gesture interpretation challenges. Their explanations, though occasionally flawed, can reveal patterns of misrecognition and model bias, informing the refinement of training data, interaction design, and future research on machine understanding of human gestures. While our study used VLM for pedestrian gesture recognition in the context of autonomous mobility, the same approach could address recognition challenges in other domains, such as wearable AR and assistive technologies. For instance, in a collaborative AR design session, a user might gesture to a colleague in the physical space, but the system mistakenly registers it as a selection gesture. A VLM could generate textual descriptions identifying potential causes, such as the gesture’s similarity to a command, insufficient differentiation between interaction zones, or background movement, thereby helping researchers refine training data and optimise recognition models for more reliable gesture-based interactions.

\subsection{Limitations}

The study is limited by the scope of the JAAD dataset, which, although diverse and richly annotated, may not fully capture the variability of pedestrian behaviour across geographic, cultural, and environmental contexts. Second, the findings are influenced by the specific version of GPT-4V used. As the model evolves with updates or enhancements, its capabilities and limitations may shift, potentially affecting the reproducibility of findings in future versions.

\begin{acks}
This research was supported by the Australian Research Council (ARC) Discovery Project DP220102019, Shared-space interactions between people and autonomous vehicles. The authors thank the anonymous reviewers for their valuable comments and suggestions, which have been incorporated into the final version.
\end{acks}

\bibliographystyle{ACM-Reference-Format}
\bibliography{references}

\appendix

\section{Scripts}
\label{sec:scripts}
To support open science, all scripts and outputs are made public:\newline  
\url{https://osf.io/prqte/?view\_only=c171d411e4604b40820e7b1eab92141c}.

\begin{itemize} 
    \item \textbf{Process Annotations}:  
    The \textit{process\_annotations.py} script parses JAAD annotation XML files (downloaded from  
    \url{https://github.com/ykotseruba/JAAD/tree/JAAD_2.0/annotations}) and classifies videos into three categories:  
    \textit{gestures}, \textit{looking}, and \textit{none}.  

    \item \textbf{Random Sampling of Videos}:  
    The \textit{random\_sampling.py} script processes the output generated by \textit{process\_annotations.py} and performs random sampling of unique videos into three categories:  23 videos with gestures, 23 videos with looking only, 23 videos with neither gestures nor looking.

    \item \textbf{Extract Video Frames}:  
    The \textit{extract\_frames.py} script uses OpenCV to extract evenly spaced frames from each video, regardless of duration. The extracted frames are saved as JPEG files within their respective video folders.

    \item \textbf{Analyse Video Frames}:  
    The \textit{analyse\_frames.py} script employs OpenAI’s Vision API to process multiple Base64-encoded images and respond to a structured prompt.  
\end{itemize}

\textbf{Notes:}
\begin{itemize}
    \item All scripts contain placeholder paths using \texttt{\$HOME} for the user’s home directory.  
    Users must manually replace these placeholders with the correct absolute paths before running the scripts.
    
    \item To use the OpenAI API, users must obtain an API key from OpenAI:  
    \url{https://platform.openai.com/api-keys}. The API key must be set in the script before execution.

    \item To run the scripts, the user must install OpenCV and OpenAI. 

\end{itemize}

\section{Examples of misrecognition and ambiguities}
\begin{figure*}[h!]
    \centering
    \includegraphics[width=0.9\linewidth]{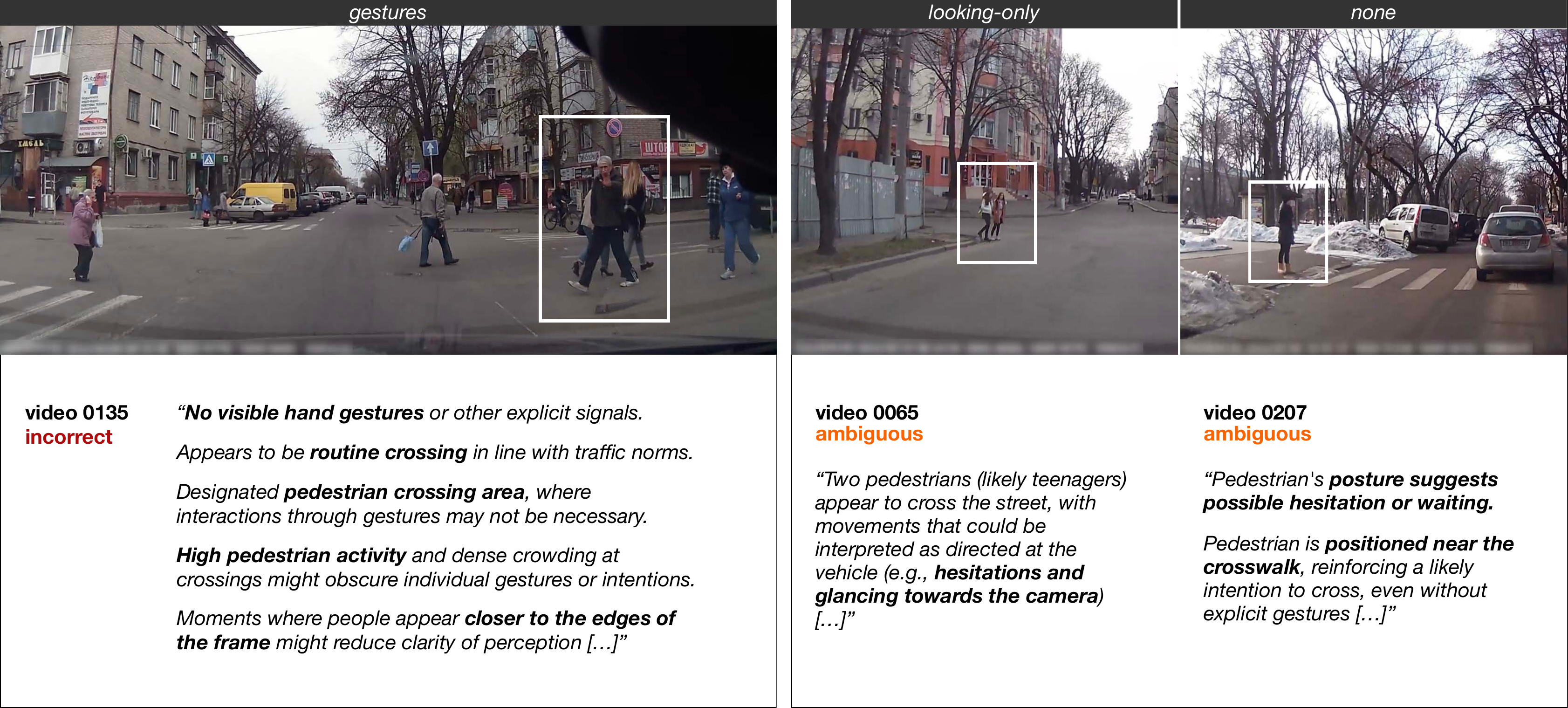}
    \caption{Examples of misrecognition and ambiguities, accompanied by excerpts from GPT-4V's reasoning. (Left) The model failed to recognise a stop gesture. (Right) The model was uncertain in two cases where no explicit gestures were present.}
    \Description{Examples of misrecognition and ambiguities, accompanied by excerpts from GPT-4V's reasoning. (Left) The model failed to recognise a stop gesture. (Right) The model was uncertain in two cases where no explicit gestures were present.}
    \label{fig:example}
\end{figure*}


\end{document}